\documentclass[12pt]{iopart}
\pdfoutput=1

\usepackage{graphicx}
\usepackage{caption}
\usepackage{subcaption}
\usepackage{epstopdf}
\usepackage{iopams}
\pdfoutput=1 
\usepackage{color}
\usepackage[numbers,square, comma, sort&compress]{natbib}
\usepackage{hyperref}

\newcommand\eq[1]{Eq.~(\ref{#1})} 
\newcommand{\eqref}[1]{Eq.~(\ref{#1})}

\usepackage{graphicx}

\def\bk{{\bf k}}
\newcommand{\beq}{\begin{eqnarray}}
\newcommand{\eeq}{\end{eqnarray}}
\def\be{\begin{equation}}
\def\ee{\end{equation}}

\begin{document}

\title[Exact results for the  Casimir force of relativistic Bose gas]{Exact results for the  Casimir force of a three-dimensional model of relativistic Bose gas in a film geometry}

\author{Daniel M Dantchev}
\address{Institute of Mechanics, Bulgarian Academy of Sciences\\
	Acad. G. Bonchev St., Building 4, 1113 Sofia, Bulgaria}
\ead{daniel@imbm.bas.bg}
	
\date{\today}
\begin{abstract}
Recently it has been suggested that relativistic Bose gas of some type can be playing a role in issues like dark matter, dark energy, and in some cosmological problems. In the current article, we investigate one known exactly solvable model of a three-dimensional statistical-mechanical model of relativistic Bose gas that takes into account the existence of both particles and antiparticles. We derive exact expressions for the behavior of the Casimir force for the system subjected to film geometry under periodic boundary conditions. We show that the Casimir force between the plates is attractive, monotonic as a function of the temperature scaling variable, with a scaling function that approaches at low temperatures a universal negative constant equal to the corresponding one for two-component three-dimensional Gaussian system. The force decays with the distance in a power-law near and below the bulk critical temperature $T_c$ of the Bose condensate and exponentially above $T_c$.  We obtain closed-form exact expression for the Casimir amplitude $\Delta_{\rm Cas}^{\rm RBG} =-4\zeta(3)/(5\pi)$.   We establish the precise correspondence of the scaling function of the free energy of the model with the scaling functions of two other well-known models of statistical mechanics - the spherical model and the imperfect Bose gas model.

\vspace{2pc}
\end{abstract}
\maketitle

\section{Introduction}

The current article is devoted to the Casimir effect in relativistic Bose gas. The role of the onset of Bose-Einstein condensation (BEC) in
a variety of limiting situations has recently been  examined in diverse fields like: relativistic superfluids and quark matter \cite{NA2005}; a relativistic pion gas \cite{BG2008}; relativistic Bose gas trapped in a generic power-law potential \cite{SuCC2006}; possible BEC of relativistic scalar field dark matter \cite{Ure_a_L_pez_2009}, see also \cite{BH2007}; relativistic Bose gas at finite chemical potential and its relation to the sign problem in QCD \cite{Aarts2009}; particle number fluctuations in relativistic Bose and Fermi gases \cite{BG2006}. The possibility that due to their superfluid properties some compact astrophysical objects may contain a significant part of their matter in the form of a Bose-Einstein condensate is envisaged in \cite{CH2012}. In Ref. \cite{FFLKT2010} the relativistic BEC is considered as a new system for analogue model of gravity. A relativistic boson gas of particles and antiparticles in the Einstein universe at high temperatures and densities is studied in \cite{PZ91}. In \cite{FM2006} a cosmological model is proposed in which dark energy is identified with
the BEC of some boson field. Let us note that according to \cite{EKJSC2018} a rapidly expanding BEC can be considered as a laboratory model of an expanding universe. The superfluidity in atomic Fermi gases and the role Bardeen-Cooper-Schrieffer (BCS) - BEC crossover plays there is reviewed in \cite{YUC2010} - see also \cite{SPRSU2018}. The above list of references is by far not exhaustive.  

In the current article we study the behavior of the Bose gas in one well known model of statistical physics of a relativistic Bose gas \cite{SP83,SP85c,PB2011}. We hope that our article clarifies some questions concerning the type of interactions between bodies immersed in a relativistic Bose condensate. We will demonstrate that if two parallel plates are immersed in such a gas then they attract each other. The force of attraction turns out to be long-ranged, i.e., to decay in a power law with the distance at and below the bulk critical temperature $T_c$ of the condensate. It is short-ranged, i.e., it decays exponentially only above $T_c$. We determine the scaling function of this force, called customary today a Casimir force, in terms of the appropriate scaling variables and determine explicitly its value, known as the Casimir amplitude, at the bulk critical point of the system. 

The Casimir effect is dubbed so after the Dutch physicist H. B. G.  Casimir. In 1948 \cite{C48}, after a discussion with Niels Bohr \cite{C99}, he realized that the zero-point fluctuations of the electromagnetic field in vacuum lead to a force of attraction between two perfectly conducting parallel plates and calculated this force.  In 1978 Fisher and De Gennes \cite{FG78} pointed out that a very similar effect exists in fluids with the fluctuating field being
the field of its order parameter, in which the interactions in the system are mediated not by photons but by different type of massless excitations such as critical fluctuations or Goldstone bosons (spin waves). Nowadays one usually terms the corresponding Casimir
effect the critical or the thermodynamic Casimir effect \cite{BDT2000}. 

Currently the Casimir, and Casimir-like, effects are object of studies in
quantum electrodynamics, quantum chromodynamics, cosmology, condensed matter physics, biology and, some elements of it, in nano-technology. The interested reader can consult the existing impressive number of reviews on the subject, see, e.g., Refs.  \cite{PMG86,MT88,LM93,MT97,M94,KG99,B99,BMM2001,M2001,M2004,L2005,BKMM2009,KMM2009,FPPRJLACCGKKKLLLLWWWMHLLAOCZ2010,OGS2011,CP2011,MAPPBE2012,B2012,WDTRRP2016,ZLP2015,K94,BDT2000,K99,G2009,TD2010,GD2011,D2012,V2015}. So far the critical Casimir effect has enjoyed only two general reviews \cite{K94,BDT2000} and few concerning specific aspects of it \cite{K99,G2009,TD2010,GD2011,D2012,V2015,MD2018}. 

The critical Casimir effect has been already directly observed, utilizing light scattering measurements, in the interaction of a colloid spherical particle with a plate \cite{HHGDB2008} both of which are immersed in a critical binary liquid mixture. Very recently, the nonadditivity of critical Casimir forces has been experimentally demonstrated in \cite{PCTBDGV2016}. Indirectly, as  a balancing force that determines the thickness of a wetting film in the vicinity of its bulk critical point, the Casimir force has been also studied in $^4$He \cite{GC99}, \cite{GSGC2006}, as well as in $^3$He--$^4$He mixtures \cite{GC2002}. In  \cite{FYP2005} and   \cite{RBM2007} measurements of the Casimir force in thin wetting films of binary liquid mixture are also performed. The studies in the field have also enjoined a considerable theoretical attention.  Reviews on the corresponding results can be found in \cite{K99,G2009,TD2010,GD2011,D2012,V2015}.

In the recent years the topics of the Casimir effect and the corresponding Casimir force in Bose systems are gaining  attention \cite{RFPZ2005,MZ2006,GD2006,FRZ2007,DiRu2017,NJN2013,JN2013,
	NP2011,TT2017,RRP2019a,TT2019,RPR2019,
	QTP2019}. The Casimir force in ideal Bose gas with a film geometry has been studied in \cite{MZ2006,GD2006,DiRu2017}. In Ref. \cite{MZ2006} the question has been treated for the first time for the case of periodic, Dirichlet and Neumann boundary conditions. In Ref. \cite{GD2006} it has been shown that the problem can be reduced to calculating the force within the Gaussian model of the properly defined $O(n)$ models, which consideration has been performed earlier in Ref. \cite{KD92a}. Then, in Ref. \cite{DiRu2017} the studies of the ideal Bose gas have been extended to Robin boundary conditions. Explicit expressions for the scaling function of the force under periodic, antiperiodic, Dirichlet, and Neumann boundary conditions have also been provided. The case of the imperfect Bose gas, with a mean-field like interaction term, has been investigated in \cite{NJN2013,JN2013,NP2011,DiRu2017}. The main conclusion of these studies is that the bulk system is characterized by the critical exponents of the spherical model \cite{BDT2000,B82,J72} and that under periodic and Dirichlet boundary conditions \cite{DiRu2017} the model with a film geometry is equivalent to the properly defined  interacting Bose gas with $2n$ internal degrees of freedom in the limit ${n\to\infty}$, i.e., the ``spherical model limit''. In a short-hand notation, one can term this model ``$O(2n)$'' model for ${n\to\infty}$, see Ref. \cite{DiRu2017}. Then, according to the universality hypothesis \cite{K71}, all these models are expected to possess the same scaling function of the free energy and the Casimir force. Only the names of the quantities involved and, therefore, the corresponding physical meaning, are different. The last turns out to be indeed correct as it is demonstrated in \cite{NJN2013,JN2013,NP2011,DiRu2017}. In the current article we study the Casimir effect in relativistic Bose gas. The bulk critical behavior of the model for general space dimension $d$ has been considered in Ref. \cite{SP83}. It has been demonstrated there that the critical exponents of this model are also equal to those ones of the spherical model. Thus, one expects, on the basis of universality, that its scaling functions of the free energy and of the Casimir force in terms of properly defined scaling variables shall be equal to those of the imperfect Bose gas and of the spherical model. We will derive explicit exact results for the scaling function of the force under periodic boundary conditions and will demonstrate that the universality is obeyed. We will show that the force is attractive  in the whole region of the thermodynamic parameters considered. A closed form expression for the Casimir amplitude will be also obtained. Finally, we will discuss the precise mapping of the relativistic Bose gas model onto the mean spherical model \cite{D98} and the imperfect Bose gas \cite{NJN2013} of classical systems. Before passing to doing that let, for completeness, mention that in addition to the ideal Bose gas and the imperfect Bose gas some results  are available for the fluctuation-induced interaction between two impurities in a weakly-interacting one-dimensional Bose gas \cite{RRP2019a,RPR2019} and, more generally, in quantum liquids \citep{RFPZ2005,FRZ2007}. BEC  mixtures have been objects of study in \cite{TT2017,TT2019,QTP2019}. Furthermore, measurement of the Casimir-Polder force through center-of-mass oscillations of a BEC has been reported in Ref. \cite{HOMC2005}.

As already stated above, in the current article we study the Casimir effect in relativistic Bose gas in space dimension $d=3$. Let us start by recalling that Bose-Einstein condensation can only occur  when the particle number is conserved \cite{H87}. Thus, in any discussion of the Bose-Einstein condensation for a relativistic Bose gas composed of particles with nonzero rest mass $m$,
at temperatures such that $k_B T ={\cal O}(mc^2)$ or greater, the possibility of particle-antiparticle pair production cannot
be ignored and must be taken into account \cite{HW81,HW82,SP83,SP84a,SP84,SP85c}. Below, in Section \ref{sec:model} we will formulate the corresponding model in the way used in \cite{SP85c}.  The results obtained there contain some of the needed expressions for the free energy in a film geometry which can be used as a starting point for deriving the corresponding results for the Casimir force.  Then, in Section \ref{sec:FSB}, we derive exact results for the behavior of the scaling function of the free energy in a film geometry - see Section \ref{sec:free_energy}, excess free energy, the Casimir force and the Casimir amplitude - see Section \ref{sec:excess_fe}. The technical details needed to clarify the precise mapping of the relativistic Bose gas onto the spherical model and the imperfect Bose gas are given in \ref{ap:equivalence} - see \ref{sec:SM} and \ref{sec:SM_excess} for the spherical model, and \ref{sec:IBG} and \ref{sec:IBG_free_energy} for the imperfect Bose gas. The article closes with a Section \ref{sec:discussion}, where we discuss several points connected to the relations between the models and some issues about the effective interactions within the system which these models actually tacitly imply.

\section{The model}
\label{sec:model}

In  \cite{SP85c} the authors consider an ideal Bose gas composed of $N_1$ particles
and $N_2$ antiparticles, each of mass $m$, confined to a
three-dimensional cuboid cavity of sides $L_1, L_2$ and $L_3$ under periodic boundary conditions. Since particles and antiparticles are 
created only in pairs, the system is governed by the conservation
of the number $Q=N_1-N_2$, which
may be looked upon as a kind of generalized ``charge''. Thus, in
equilibrium, the chemical potentials of the two species are equal and opposite, i.e., $\mu_1=-\mu_2=\mu$. With respect to the occupation numbers $N_1$ and $N_2$ this results in 
\be
\label{eq:N1_N2_Bose}
N_1=\sum_{\varepsilon(\bk)}\left[e^{\;\beta(\varepsilon-\mu)}-1\right]^{-1}, N_2=\sum_{\varepsilon(\bk)}\left[e^{\;\beta(\varepsilon+\mu)}-1\right]^{-1},
\ee
where 
\be
\label{eq:energy}
\varepsilon(\bk)=\sqrt{{\bk}^2+m^2}.
\ee
We are using in  \eq{eq:N1_N2_Bose}, \eq{eq:energy}, and thereafter, the units $\hbar=c=k_B=1$, thus $\beta=1/T$, and let $L_i, i=1,2,3$ are measured in terms of some microscopic length scale, i.e., $L_i, i=1,2,3$ are dimensionless. Then, under periodic boundary conditions, the eigenvalues $k_i, (i= 1,2, 3)$ of the wave vector $\mathbf{k}$ are given by
$k_i =(2\pi/L_i)\,n_i$, where $n_i =0, \pm 1, \pm 2, . . . $. 
Let us stress that here  both $\varepsilon$ and $\mu$ include the rest energy $m$ of the particle, or of the antiparticle.
The condition $|\mu|\le m$ ensures that the mean occupation numbers in the various
states are positive definite. Obviously, one has two symmetric cases $\mu>0$ and $\mu<0$. If, for definiteness, one assumes $\mu>0$ it follows that $Q>0$, i.e., $N_1>N_2$. In view of conservation of $Q$, $\mu$ shall keep its sign.  Thus, for definiteness in what follows we assume $\mu>0$. 

\section{On the finite size behavior of the model in film geometry}
\label{sec:FSB}

The pressure $P$ in the grand canonical ensemble \cite{SP85c,PS2016} may then be written as 
\begin{equation}
\label{eq:gran_pot_P}
P=-\frac{1}{\beta V} \sum_{\varepsilon_{\mathbf n}} \left[\ln\left(1-e^{-\beta(\varepsilon_{\mathbf n}-\mu)}\right)\right]+\left[\ln\left(1-e^{-\beta(\varepsilon_{\mathbf n}+\mu)}\right)\right], 
\end{equation}
where ${\mathbf n}=\{n_1,n_2,n_3\}$, and $V=L_1 L_2 L_3$. In accord with the standard thermodynamic relation, for the charge density one has
\begin{equation}
\label{eq:cgarge_density}
\rho\equiv \frac{Q}{V}=\left(\frac{\partial P}{\partial \mu}\right)_{T}.
\end{equation}

	Using the identity
	\begin{equation}
	\label{eq:identity}
	\sum _{j=1}^{\infty } \frac{\cosh (j a) \exp (-j
		b)}{j}=-\frac{1}{2} \left \{\log \left(1-e^{-(b-a)}\right)+\log
	\left(1-e^{-(b+a)}\right)\right \},
	\end{equation}
expression in \eq{eq:gran_pot_P} can be reorganized in the more convenient form
\begin{eqnarray}
\label{eq:P-new_repres}
&& P(\beta,\mu,m|L_1,L_2,L_3)=\frac{2}{\beta V}\sum_{j=1}^{\infty}\frac{\cosh (j \beta \mu)}{j}\\ && \sum_{n_1=-\infty}^{\infty}\sum_{n_2=-\infty}^{\infty}\sum_{n_3=-\infty}^{\infty}\exp\left[-j\beta m \sqrt{1+\frac{4\pi^2}{m^2}\sum_{i=1}^{3} \left(\frac{n_i}{L_i}\right)^2}\,\right]. \nonumber
\end{eqnarray}

In Refs. \cite{SP84a} and \cite{SP84} specific techniques for dealing with sums of the above type have been developed. With their help, in Ref. \cite{SP85c} some results for the scaling function of the free energy in {\it a)} fully finite, {\it b)} square channel  and {\it c)} film geometry have been reported. More specifically, one considered three dimensional systems with periodic boundary conditions in geometry of a {\it a)} cube, i.e., $L_1=L_2=L_3=L$, {\it b)}  square channel, i.e.,  $L_1\to\infty$, $L_2=L_3=L$, and {\it c)} a film, i.e., $L_1,L_2\to\infty$, $L_3=L$.  In the current article we are mainly interested in a system with a film geometry with a finite thickness $L$.   For this case in Ref. \cite{SP85c}  explicit results only for the low and high-temperature asymptotic of the scaling function are  presented. For the intermediate region $L(T-T_c)/T_c=O(1)$ just a numerical evaluation of the scaling function at the bulk critical point $T=T_c$ is given \cite[p. 1822]{SP85c}. In the current article we will filly cover this region obtaining explicit results for the behavior of the free energy, the Casimir force and the Casimir amplitude. 

The most general expressions are, of course, pertinent to the fully finite system. The remaining ones can be obtained by taking the appropriate limits, as specified above. In order to be specific, and to introduce the  notations needed, let us present these expressions \cite{SP85c}
\be
\label{eq:fully_finite}
P=\frac{m^4}{2\pi^2}X(\beta,\mu)+\frac{1}{2\pi \beta}\left[\sqrt{m^2-\mu^2}H_2(\mu)+H_3(\mu) \right],
\ee
where
\begin{equation}
\label{eq:X}
X(\beta,\mu)=2\sum_{j=1}^\infty \cosh(j\beta\mu) \frac{K_2(j\beta m)}{(j \beta m)^2},
\end{equation}
and 
\begin{equation}
\label{eq:Hn}
H_n(\mu)=\sum_{\mathbf{q}}\!'  \frac{\exp\left[-\sqrt{m^2-\mu^2}\gamma(\mathbf{q})\;\right]}{\gamma^n(\mathbf{q})},
\end{equation}
with
\begin{equation}
\label{eq:gamma}
\gamma(\mathbf{q})=\sqrt{\sum_{i=1}^3 q_i^2L_i^2}, \quad  \mbox{where} \quad  \mathbf{q}=\{q_1, q_2, q_3\}, \quad q_i =0, \pm 1, \pm 2, . . . .
\end{equation}
In \eq{eq:X} $K_2$ is the modified Bessel function of the second kind. In \eq{eq:Hn} the prime means  that the term with $\mathbf{q}=\mathbf{0}$ is omitted and, therefore, $\gamma(\mathbf{q})>0$.

\subsection{On the finite size behavior of the free energy}
\label{sec:free_energy}

Obviously, in \eq{eq:fully_finite} the function $X(\beta,\mu)$ reflects the bulk behavior, while the terms in the quadratic brackets take into  account the effects related to the finite extensions of the system. 
Taking now the limits $L_1,L_2\to\infty$ and setting $L_3=L$, we obtain the corresponding basic result for the film geometry. Eqs.  (\ref{eq:fully_finite}) and (\ref{eq:X}) stay formally the same, only \eq{eq:Hn} simplifies to 
\begin{eqnarray}
\label{eq:Hn_film}
H_n(\mu)&=&2\sum_{q=1}^\infty \frac{\exp\left[-q\sqrt{m^2-\mu^2}L \right]}{(q L)^n}\\
&=&2\left(m^2-\mu^2\right)^{n/2}(2y_L)^{-n}\mathrm{Li}_n\left(e^{-2
	y_L}\right), \nonumber
\end{eqnarray}
where we have introduced the parameter
\beq
\label{eq:y_L_def}
y_L=\frac{1}{2}\sqrt{m^2-\mu^2}L.
\eeq
Then, \eq{eq:fully_finite} becomes
\begin{equation}
\label{eq:scaling}
P=\frac{m^4}{2\pi^2}X+L^{-3}\frac{1}{\pi  \beta }\left[2 y_L\,
	{\rm Li}_2\left(e^{-2 y_L}\right)+{\rm Li}_3\left(e^{-2
		y_L}\right)\right]. 
\end{equation}
In \eq{eq:Hn_film} $\mathrm{Li}_n(z)$ is the polylogarithm function, also known as the Jonquière's function
\begin{equation}
\label{eq:Li-def}
\mathrm{Li}_n(z)=\sum_{k=1}^\infty z^k/k^n.
\end{equation}
The $\mathrm{Li}_n(z)$ are directly related to the  Bose-Einstein functions \cite{PB2011}
\begin{equation}
\label{eq:BE-function}
g_\nu(z)=\frac{1}{\Gamma(\nu)}\int_{0}^{\infty}
\frac{x^{\nu-1}dx}{z^{-1}e^x-1}.
\end{equation}
It is easy to show \cite{PB2011} that $\mathrm{Li}_\nu(z)=g_\nu(z), 0\le z<1$. Let us also note that sometimes $\mathrm{Li}_n(z)$ are denoted as $F(z,n)$, or $F_n(z)$ \cite{GB68} functions. Due to the above mentioned diversity in notations one can encounter results for the Bose gas formulated in terms of different but otherwise equivalent functions. In the current article we will use formulations in terms of polylogarithm functions  $\mathrm{Li}_n(z)$. As we will see later, technically this is an important moment because the available identities for these functions will allow us to obtain closed form explicit solution for the Casimir amplitude of the model. 

From \eq{eq:cgarge_density} and \eq{eq:scaling}, in the case of a film geometry one has 
\begin{eqnarray}\label{eq:rho_film}
\lefteqn{\rho = \frac{m^3}{2\pi^2} W(\beta,\mu)-\frac{\mu}{2\pi\beta}\sqrt{m^2-\mu ^2} \frac{\log \left(1-e^{-2 y_L}\right)}{y_L}} \nonumber\\
&=&\frac{m^3}{2\pi^2} W(\beta,\mu)+ \frac{1}{L} \frac{\mu}{\pi\beta}
\left[y_L-\log (2 \sinh
	y_L)\right]. \;\;\;\;\;\;\;\;\;
\end{eqnarray}
Here 
\begin{equation}
\label{eq:W_function}
W(\beta,\mu) = m \left(\frac{\partial X}{\partial \mu} \right)_\beta=2\sum_{j=1}^\infty \sinh(j\beta\mu) \frac{K_2(j\beta m)}{j \beta m}.
\end{equation}

From \eq{eq:scaling} and \eq{eq:rho_film} for the ``thermal'', see Ref.  \cite{SP85c},  free energy density   of the system, i.e., the part of the free energy that is temperature dependent, one obtains
\begin{eqnarray}
\label{eq:thermal_free_energy}
\bar{f}&\equiv& \frac{\bar{F}}{V}\equiv \frac{F-mQ}{V}=(\mu-m)\rho -P =-\frac{m^4}{2\pi^2}\left[X(\beta,\mu)+\frac{m-\mu}{m}W(\beta,\mu)\right] \nonumber \\
&& -L^{-3}\frac{1}{\pi  \beta }\left[2 y_L\,
{\rm Li}_2\left(e^{-2 y_L}\right)+{\rm Li}_3\left(e^{-2
	y_L}\right)\right] \nonumber \\
&&-  \frac{1}{L} \frac{\mu(m-\mu)}{\pi\beta} \left[y_L-\log (2 \sinh y_L)\right].
\end{eqnarray}
When $\mu\simeq m$ the following expansions are valid 
\begin{equation}\label{eq:W_expansion}
W(\beta,\mu)=W(\beta,m)-\frac{\pi\mu}{\beta m^3}\sqrt{m^2-\mu^2}+{\cal O}(m^2-\mu^2),
\end{equation}
and 
\begin{eqnarray}\label{eq:X_expansion}
X(\beta,\mu)&=& X(\beta,m)-\frac{m-\mu}{m}W(\beta,m)\\
&& +\frac{\pi}{3\beta m^4}(m^2-\mu^2)^{3/2}+{\cal O}(m^2-\mu^2)^2\nonumber
\end{eqnarray}
or, in terms of $y_L$,
\begin{equation}\label{eq:W_expansion}
W(\beta,\mu)=W(\beta,m)-\frac{1}{L}\frac{2\pi\mu}{\beta m^3}\, y_L+{\cal O}(y_L^2),\;\;\;\;\;
\end{equation}
and 
\begin{equation}\label{eq:X_expansion}
X(\beta,\mu)= X(\beta,m)-\frac{m-\mu}{m}W(\beta,m) +\frac{1}{L^3}\frac{8 \pi}{3\beta m^4}y_L^3+{\cal O}\left(y_L^4\right).
\end{equation}
With their help, \eq{eq:rho_film} and \eq{eq:thermal_free_energy} become
\begin{equation}\label{eq:rho_film_expanded}
\rho = \frac{m^3}{2\pi^2} W(\beta,m)- \frac{1}{L} \frac{\mu}{\pi\beta} \log (2 \sinh y_L),
\end{equation}
and
\begin{eqnarray}\label{eq:free_energy_finite_scaling}
\bar{f}&=&-\frac{m^4}{2\pi^2} X(\beta,m)\\
&& +L^{-3}\frac{1}{\pi  \beta }\bigg\{\frac{2}{3}y_L^3-2 y_L\,
{\rm Li}_2\left(e^{-2 y_L}\right)-{\rm Li}_3\left(e^{-2
	y_L}\right) \nonumber\\
&&  -2y_L^2\left[y_L-\log (2 \sinh y_L)\right] \bigg\}. \nonumber
\end{eqnarray}
Thus, having in mind that $X(\beta,m)$ is a regular term, in a full accord with Ref. \cite{SP85c} the singular part, i.e., the part possessing scaling behavior, of the free-energy density is 
\begin{eqnarray}\label{eq:sing_part_free_energy}
&&f^{(s)}(T;L)\equiv \frac{\beta \bar{F}^{(s)}}{V} \\
&=&L^{-3}\frac{1}{\pi }\bigg\{\frac{2}{3}y_L^3-2 y_L\,
{\rm Li}_2\left(e^{-2 y_L}\right)-{\rm Li}_3\left(e^{-2
	y_L}\right) \nonumber\\
&&  -2y_L^2\left[y_L-\log (2 \sinh y_L)\right] \bigg\}.\nonumber
\end{eqnarray}
Recalling now that the bulk critical point $\beta_c$ is determined by the condition
\begin{equation}
\label{eq:crit_point}
\rho=\frac{m^3}{2\pi^2} W(\beta_c,m),
\end{equation}
see \eq{eq:rho_film} with $\mu(\beta_c)=m$,
\eq{eq:rho_film_expanded} becomes 
\begin{equation}\label{eq:rho_film_expanded_what_is_t}
W(\beta,m)-W(\beta_c,m)=\frac{1}{L} \frac{2\pi }{\beta_c m^2} \log (2 \sinh y_L).
\end{equation}
Expanding the above about $\beta_c$ and introducing the notation $x_\tau$, we obtain
\begin{eqnarray}\label{eq:x_tau}
x_\tau&\equiv&\beta_c m^2 L
\left[  
W(\beta,m)-W(\beta_c,m)
\right] \\
&&\simeq\left( \beta_c^2 m^2\left|\frac{\partial W}{\partial \beta}\right|_{\beta=\beta_c}\right)\, L\tau,\; \quad \mbox{with} \quad \tau=\frac{T-T_c}{T_c},\nonumber
\end{eqnarray}
i.e., \eq{eq:rho_film_expanded_what_is_t} now reads 
\begin{equation}\label{eq_x_tau_q_L}
x_\tau=2\pi \log (2 \sinh y_L).
\end{equation}
Thus, for the singular part of the free energy density one has
\begin{equation}\label{eq:free_energy_density_scaling}
\bar{f}=L^{-3}X_f(x_\tau),
\end{equation}
where $X_f(x_\tau)$ is determined, see \eq{eq:sing_part_free_energy}, by the expression 
\begin{eqnarray}\label{eq:scaling_function}
X_f(y_L) &=& \frac{1}{\pi }\bigg\{\frac{2}{3}y_L^3-2 y_L\,
{\rm Li}_2\left(e^{-2 y_L}\right)-{\rm Li}_3\left(e^{-2
	y_L}\right) \nonumber\\
&&  -2y_L^2\left[y_L-\log (2 \sinh y_L)\right] \bigg\}.
\end{eqnarray}
where $y_L$, according to \eq{eq_x_tau_q_L}, is 
\begin{equation}\label{eq:y_L_via_x_tau}
y_L(x_\tau)={\rm arcsinh}\left[\frac{1}{2}\exp\left(\frac{x_\tau}{2\pi} \right)\right].
\end{equation}
It is easy to check that $y_L(x_\tau)$ is a monotonically increasing function of $x_\tau$. Then, since
\begin{equation}\label{eq:der_Xf_x_tau}
\frac{d X_f[y_L(x_\tau)]}{d x_\tau}=\frac{1}{\pi ^2}\rm{csch}^{-1}\left[2 \exp{\left(-\frac{x_\tau}{2 \pi
	}\right)}\right]^2,
\end{equation}
we conclude that $X_f$ is, as it is to be expected, a monotonically increasing function of $x_\tau$. 

In Ref. \cite{SP85c} the only specific results reported for $X_f$, are those ones of the values of $X_f$, see Eq. (67) there, for low temperatures, i.e., for $t<0$ when $L\to \infty$, and in the opposite case of high temperatures, i.e., when $t>0$ and $L\to\infty$. One finds that 
\be
X_f(x_\tau)\simeq \frac{1}{12 \pi^4} x_\tau^3,\quad x_\tau\to \infty
\label{eq:Xf_Bose_above}
\ee
which is, in fact, the bulk result $X_f^{(b)}(x_\tau)$, and
\be
\label{eq:Xf_Bose_below}
X_f(x_\tau)=-\frac{1}{ \pi} \zeta(3), \quad x_\tau\to -\infty.
\ee
These results are obviously easily reproducible from \eq{eq:scaling_function} and \eq{eq:y_L_via_x_tau}, taking into account that $y_L\to\infty$ when $x_\tau\to\infty$, and $y_L\to 0$ when $x_\tau\to-\infty$. Furthermore, from \eq{eq:y_L_via_x_tau} with $x_\tau=0$ one gets $y_L(x_\tau=0)={\rm arcsinh}[1/2]\simeq 0.481212$ which is basically the value of $0.48$ reported in \cite[p. 1822]{SP85c}.

\subsection{On the behavior of the excess free energy and the Casimir force }
\label{sec:excess_fe}

From \eq{eq:free_energy_density_scaling}, \eq{eq:scaling_function}, and  \eq{eq:Xf_Bose_above} it is easy to obtain the excess free energy normalized per unit area
\begin{equation}\label{f_ex}
\beta f_{\rm ex}(x_\tau)=L^{-(d-1)} X_{\rm ex}(x_\tau), \quad \mbox{with} \quad d=3,
\end{equation}
where
\begin{equation}\label{Xex_def}
X_{\rm ex}(x_\tau)= X_f(x_\tau)-X_f^{(b)}(x_\tau),
\end{equation}
i.e., the amount of free energy in excess to the bulk one. Explicitly, one has 
\beq
\label{eq:Xex_Bose}
X_{{\rm ex}}(x_\tau)&=&\frac{1}{\pi}\Bigg[\frac{2 }{3  }\left(y_L^3-y_\infty^3\right)+2 y_L^2\ln \left(1-e^{-2 y_L}\right) \nonumber\\
&& -2 y_L
\mathrm{Li}_2\left(e^{-2 y_L}\right)-\mathrm{Li}_3\left(e^{-2
	y_L}\right)\Bigg],
\eeq
where $y_L$ is given by \eq{eq:y_L_via_x_tau}, and 
\begin{equation}
\label{eq:yi_Bose}
 y_\infty=\left\{\begin{array}{ll}
 	x_\tau/(2 \pi), & x_\tau\ge 0 \\
 	0, & x_\tau \le 0.
 \end{array}\right.
\end{equation}

From the excess free energy we can derive the corresponding expression for the Casimir force. By definition \cite{BDT2000}
\begin{equation}\label{key}
\beta F_{\rm Cas}=-\frac{\partial\left[ \beta f_{\rm ex}(x_\tau)\right]}{\partial L}.
\end{equation}
From the above definitions it follows that 
\begin{equation}\label{eq:Casimir_force}
\beta F_{\rm Cas}=L^{-d}X_{\rm Cas}(x_\tau), \quad d=3,
\end{equation}
where the scaling function of the Casimir force is
\begin{equation}\label{eq:scaling_function_excess_free_energy}
X_{\rm Cas}(x_\tau)=(d-1)X_{\rm ex}(x_\tau)-\frac{1}{\nu}x_\tau\frac{\partial X_{\rm ex}(x_\tau)}{\partial x_\tau}.
\end{equation} 
In the system considered here $d=3$. From \eq{eq:x_tau}, taking into account that, according to the general theory \cite{BDT2000}, one shall have  $x_\tau=C tL^{1/\nu}$ with $C$ being a given system dependent constant, and $\nu$ the critical exponent of the correlation length, we conclude that $\nu=1$. Performing the calculations, from \eq{eq:scaling_function_excess_free_energy} one derives 
\begin{eqnarray}\label{eq:XCas_final}
X_{\rm Cas}(x_\tau)&=&-\frac{2}{\pi} \bigg [\frac{1}{3}
\left(y_L^3-y_\infty^3\right)+2 y_L {\rm Li}_2\left(e^{-2
		y_L}\right)\\
	&& +{\rm Li}_3\left(e^{-2 y_L}\right)-y_L^2 \log
	\left(1-e^{-2 y_L}\right)\bigg ].\;\;\;\;\;\; \nonumber
\end{eqnarray}
It is easy to check that $y_L\ge y_\infty$ and, thus, all terms in quadratic brackets are positive, i.e. $X_{\rm Cas}(x_\tau)\le 0$. The last implies that within the relativistic Bose gas the Casimir force is always \textit{attractive}.  

\begin{figure}[h!]
	\includegraphics[width=\columnwidth]{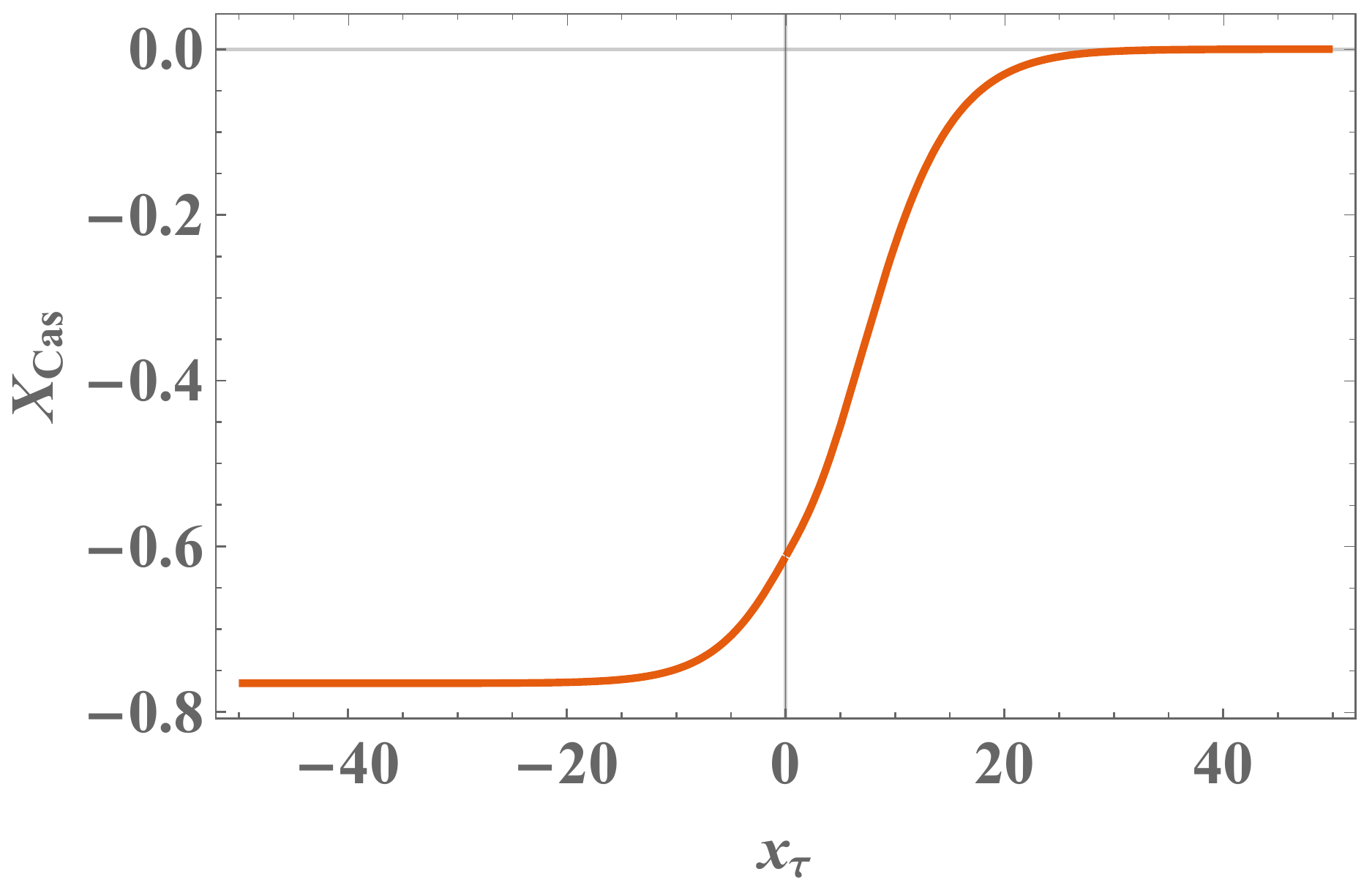}
	\caption{ The behavior of the Casimir force within the relativistic Bose gas. We observe that the scaling function is negative, monotonic, and approaches an universal negative constant for low temperatures. }
	\label{fig:PsiSRPR}
\end{figure}

The Casimir amplitude can be derived from \eq{eq:Xex_Bose}, or \eq{eq:XCas_final}. From \eq{eq:yi_Bose} one has $y_\infty(x_\tau=0)=0$ and from \eq{eq:y_L_via_x_tau} that $y_L(x_\tau=0)={\rm arcsinh}(1/2)$. Plugging these values in, say, \eq{eq:Xex_Bose}, after some manipulations based on identities for the polylogarithmic functions, see below, one obtains  
\begin{equation}\label{eq:Cas_ampl_RBG}
X_{\rm ex}(x_\tau=0)=\frac{1}{2}X_{\rm Cas}(x_\tau=0) \equiv \Delta_{\rm Cas}^{\rm RBG} =-\frac{4}{5\pi}\zeta(3).
\end{equation}
Let us briefly elucidate the procedure that leads to the above explicit result. First, let us note that 
\begin{equation}\label{eq:identities}
{\rm arcsinh}\left(\frac{1}{2}\right)=\log \left(\frac{1}{2}+\frac{\sqrt{5}}{2}\right)\equiv \log \varphi,
\end{equation}
where $\varphi$ is the so-called golden ratio
\begin{equation}\label{eq:gr}
\varphi=\frac{1}{2}+\frac{\sqrt{5}}{2}.
\end{equation}
Then
\begin{equation}\label{eq:via_gr}
\exp[-2y_L(x_\tau=0)]=\varphi^{-2}=2-\varphi,
\end{equation}
and \eq{eq:Xex_Bose} takes the form 
\beq
\label{eq:Xex_Bose_crit_point}
X_{{\rm ex}}(x_\tau=0)&=&\frac{1}{\pi}\Bigg[\frac{1 }{6 }\log^{3} (2-\varphi)\\
&& + \log (2-\varphi) \,
\mathrm{Li}_2\left(2-\varphi\right)-\mathrm{Li}_3\left(2-\varphi\right)\Bigg], \nonumber
\eeq
i.e., it can be written entirely in terms of the golden ratio. Using now the identity \cite{L81}, see also \cite{Sa93}
\beq
\label{eq:identity_polylog} 
\log (2-\varphi) \,
\mathrm{Li}_2\left(2-\varphi\right)-\mathrm{Li}_3\left(2-\varphi\right)=-\frac{1 }{6 }\log^{3} (2-\varphi)-\frac{4\zeta(3)}{5},
\eeq
we arrive at the result reported in \eq{eq:Cas_ampl_RBG}. Actually, it can be shown \cite{L81,Sa93} that at $z=2-\varphi$  both $\mathrm{Li}_3(z)$ and $\mathrm{Li}_2(z)$ can be expressed in terms of elementary functions
\beq
\label{eq:identity_polylog_Levin} 
\mathrm{Li}_3\left(2-\varphi\right) &=&\frac{4}{5}\zeta(3)+\frac{1}{15}\pi^2 \log(2-\varphi)-\frac{1}{12}\log^3(2-\varphi), \nonumber \\
\mathrm{Li}_2\left(2-\varphi\right) &=& \frac{1}{15} \pi^2-\frac{1}{4} \log^2(2-\varphi).
\eeq

\section{Concluding remarks and discussion}
\label{sec:discussion}

In the current article we have derived exact analytical expressions for the singular part of the free energy, see \eq{eq:scaling_function}, excess free energy scaling function, see \eq{eq:Xex_Bose}, and the Casimir force, see \eq{eq:XCas_final}, for the relativistic Bose gas. We have determined explicit expression for the Casimir amplitude of the model, see \eq{eq:Cas_ampl_RBG}, and have shown that the Casimir force is a monotonically increasing with temperature {\it attractive} force, see Fig.  \ref{fig:PsiSRPR}. We have also determined the low temperature asymptotic of the free energy, i.e., of the excess free energy, see \eq{eq:Xf_Bose_above}. 
Let us recall that the Casimir amplitude \cite{KD92a} for the $n$-component Gaussian model with  $d\in (2,4)$  is
\begin{equation}
\label{eq:Cas_Gaussian}
\Delta_{\rm Cas}^{\rm GM}(n,d)=-n \frac{\Gamma(d/2)}{\pi^{d/2}} \zeta(d).
\end{equation}
With $n=2$ and $d=3$ this leads to $\Delta_{\rm Cas}^{\rm GM}(2,3)=-\zeta(3)/\pi$.
Thus, the low temperature asymptotic of the finite-size part of the free energy is equal to that one of the two-component Gaussian model. 

The model considered here is characterized with $d=3$ and, as we have seen, $\nu=1$. As already stated in the introduction, the bulk critical behavior of the model for general $d$ has been considered in \cite{SP83}. For  $d\in (2,4)$ one has that 
\begin{equation}
\label{eq:crit_exponents}
\nu=\frac{1}{d-2},\; \beta=\frac{1}{2}, \; \eta=0,\; \alpha=\frac{d-4}{d-2}, \; \gamma=\frac{2}{d-2}.
\end{equation}
These critical exponents coincide with those of the spherical model \cite{BDT2000,B82,J72} and the imperfect non-relativistic Bose gas. Then, according to the universality hypothesis \cite{K71}, all these models shall possess the same scaling function of the free energy and the Casimir force with only the names of the quantities involved and, therefore, the corresponding physical meaning, being different. Inspecting the result presented in \eq{eq:Cas_ampl_RBG}, and the corresponding one for the spherical model, see Eq. (30) in \cite{D98}, shows that the Casimir amplitude for the relativistic Bose gas is exactly two times larger than the one for the spherical model. A careful comparison, see  \ref{ap:equivalence}, of the expressions for the scaling variables and the excess free energy scaling functions shows that a mapping of one into the other is possible with the result that
\be
2X_{{\rm ex}}^{\left(\rm SM \right)}=X_{{\rm ex}}^{\left(\rm Bose\right)},
\label{eq:SM_Bose}
\ee
where the upper-scripts indicate to which model the corresponding scaling function belongs. In the same Appendix the result for the scaling function of the excess free energy $X_{\rm ex}^{(\rm IB)}$ for $T<T_c(\mu)$ for the imperfect Bose gas with a mean-field like interaction term, is also presented. The comparison with the results for the spherical model shows, again, that 
\be
\label{eq:rel_Imp_Bose_SM}
X_{\rm ex}^{(\rm IB)}=2 X_{\rm ex}^{(\rm SM)}.
\ee
Thus, for $T<T_c$, where $T_c$ is the critical temperature for the corresponding model,
\begin{equation}\label{eq:rel_Imp_Bose_Rel_Bose_gas}
X_{{\rm ex}}^{\left(\rm Bose\right)}=X_{\rm ex}^{(\rm IB)}.
\end{equation}

One might wonder how a model of ideal relativistic Bose gas is mathematically equivalent to the spherical model with short-range nearest neighbor interaction and to the imperfect Bose gas. The last implies, indeed, that the model considered is not really “purely non-interacting” one. The inspection of the conditions imposed on the model lead to the conclusion that some sort of an effective interaction comes through the requirement that the density of “charge” is fixed.  This ``interaction'' obviously acts uniformly over all particles in the system. 

Let us note that in the relativistic Bose gas, the imperfect Bose gas, and the spherical model the value of the critical temperature $\beta_c$, and the corresponding temperature dependence of the free energy about it follows from one self-consistent equation --- the equation for having a fixed length of the spins in the spherical model (“spherical field” equation - see Eq. (14) in Ref. \cite{D96}), the “stationary-point equation” (see Eq. (9) in Ref. \cite{NJN2013}) for the imperfect Bose gas, or for the charge density in the case of relativistic Bose gas (see \eq{eq:rho_film} above).  Note that when fixing the charge density $\rho$ one does not determine $\mu(\rho)$, as in the usual ensemble transformation, but $\beta_c(\rho,m)$.  This additional self-consistent equation is what makes the above-mentioned models belonging to the universality class of the $O(n)$ models in the case of $n\to\infty$, and not to the Gaussian model. This is also the main difference with the ideal Bose gas model behavior considered in Ref. \cite{GD2006} -- it is equivalent to the Gaussian model description within the $O(n)$ models’ formulation. To make the story simple: the spherical model, the imperfect Bose gas and the relativistic model are equivalent to the Gaussian model in which the parameters satisfy one self-consistent equation. Actually, all mathematical difficulty is  normally in solving this equation. The fact that the models are equivalent to the Gaussian model with one additional equation that has to be satisfied, leads to the result  that the scaling function of the free energy formally looks like the one of the Gaussian model (compare, e.g., \eq{eq:scaling} with Eq. (3.38) for $d=3$ in \cite{DK2004}; Eq. (17) in \cite{D98}; Eq. (3.10a) in \cite{DiRu2017}, or with what follows from Eq. (2) in \cite{GD2006}), but the meaning of the parameters is different -- they have to satisfy that additional equation. The last leads to critical exponents and Casimir amplitudes, and temperature dependence of the scaling functions, different from that of the Gaussian model and makes the model having well defined free energy also \textit{below} $T_c$.

The consideration presented in the current article are for periodic boundary conditions. One might expect that Dirichlet-Dirichlet boundary conditions shall be much more realistic. If the analogy with the spherical model is then further preserved, which seems plausible, one shall expect that the Casimir force is again attractive. It will be, however, no longer monotonic as a function of the temperature scaling variable but will possess a deep minimum below the bulk critical temperature - see Refs. \cite{DGHHRS2012,DBR2014}.

 Finally, let us also make some comments of how the type of dispersion relation changes the properties of the considered system. In Ref. \cite{GB68}  one studied Bose-Einstein condensation with single-particle energy spectrum $\varepsilon(\bk)\sim |{\bf k}|^\sigma$, with $0<\sigma \leq 2$. The result  is that systems with $\sigma =1$ and $\sigma =2$ belong to {\it different} universality classes for given values of dimensionality $d$ of the system. Only when the spectrum of the relativistic system, as explained in Ref. \cite{SP83,PB2011}, is with the form considered in the current article, and, as suggested by Haber and Weldon in Ref. \cite{HW81,HW82} the possibility of particle-antiparticle pair production in the system is taken into account, the relativistic and non-relativistic Bose gasses do belong to the {\it same} universality class.  The effect of changes in dispersion relation on the value of the critical temperature, which is a non-universal quantity and thus model dependent,  and the relation of $T_c$ with the number density, is studied in Ref. \cite{GLB2007}. The fact that the change of spectrum from $\varepsilon(\bk)\sim |{\bf k}|$ to $\varepsilon(\bk)\sim |{\bf k}|^2$ leads to different universality classes can be easily understood if one takes into account the mapping onto the spherical model. While the  $\varepsilon(\bk)\sim |{\bf k}|^2$ spectrum corresponds to a spherical model with short-ranged interaction, the one with $\varepsilon(\bk)\sim |{\bf k}|^\sigma$, $0<\sigma<2$ is a model with a power-law decaying interaction decaying as $r^{-d-\sigma}$ with the distance \cite{BDT2000}. The critical exponents in the last case do continuously depend on $\sigma$ for $\sigma<d<2\sigma$ \cite{PB2011,BDT2000}. For $d>2\sigma$ the system is characterized by the mean-field critical exponents. Obviously the system with $d=3$ and $\sigma=1$ do belong to the last case. For the critical behavior of the system, as it is well known, the wavelength asymptotic of the spectrum is the important one, i.e., the limit $|{\bf k}|\to 0$. Thus, for any fixed $m$ the expansion of the spectrum given by \eq{eq:energy} will lead to short-ranged type universality class.  

In the introduction of the current article we have discussed several topics in which BEC of relativistic Bose gas is of an essential importance. We hope that our considerations would be of use in some of them.

\section*{Acknowledgment}

The author gratefully acknowledges the financial support via contract DN 02/8 with Bulgarian National Science Fund.

\appendix

\section{On the relation between the relativistic Bose gas model with two other models}
\label{ap:equivalence}

In the current appendix, we demonstrate that there exists a simple relation between the excess free energy scaling functions of the relativistic Bose gas model with these ones of the spherical model and of the imperfect Bose gas with mean-field type interaction. In order to introduce the notations, we will first briefly remind the definitions of the mean spherical model and the imperfect Bose gas and will, by using results reported in the literature, demonstrate that the scaling functions of the free energy in all three models indeed coincide up to a factor of 2 in the case of relativistic and the imperfect Bose gas. Actually, in Ref. \cite{DiRu2017} one showed the equivalence of the imperfect Bose gas with an interacting Bose gas with $2n$ internal degrees of freedom in the limit ${n\to\infty}$, i.e., the ``spherical model limit''. In a short-hand notation, one can term this model ``$O(2n)$'' model for ${n\to\infty}$, see Ref. \cite{DiRu2017}. Let us note that the standard spherical model involves only a physically reasonable short-ranged pair potential.

\subsection{A short definition of the spherical model}
\label{sec:SM}

We consider a  model embedded on a $d$-dimensional hypercubic
lattice ${\cal L} \in \mathbb{Z}^d$, where ${\cal L}=L_1\times L_2\times
\cdots L_d$. Let $L_i=N_i a_i, i=1,\cdots,d$, where $N_i$ is the
number of spins and $a_i$ is the lattice constant along the axis
$i$ with ${\bf e}_i$ being a unit vector along that axis, i.e., ${\bf e}_i.{\bf e}_j=\delta_{ij}$.
With each lattice site $\bf{r}$ one associates a real-valued spin
variable $S_{\bf r}\in \mathbb{R}$ which obeys the constraint
\begin{equation}
\label{constraint}
\langle S_{\bf r}^2
\rangle=1, \qquad \rm{for\; all} \qquad {\bf r} \in {\cal L}.
\end{equation}
The average in \eq{constraint} is with respect to the
Hamiltonian of the model
\begin{equation}
\label{Hamsm}
\beta{\cal H}=-\frac{1}{2}\beta\sum_{{\bf r}, {\bf
		r}'}S_{\bf r}J({\bf r}, {\bf r}')S_{{\bf r}'}-\sum_{{\bf
		r}}h_{\bf r}S_{\bf r}+\sum_{{\bf
		r}}\lambda_{\bf r}\left(S_{{\bf r}}^2-1\right),
\end{equation}
where the Lagrange multipliers $\lambda_{\bf r}$, called {\it spherical fields}, are determined so that \eq{constraint} is fulfilled for all ${\bf r} \in {\cal L}$. Eqs. \eqref{constraint} and \eqref{Hamsm} represent the most general definition of the so-called {\it mean spherical model} \cite{BK52,LW52,K73}. Its main difference from the standard Ising model is that \eq{constraint} is fulfilled only in {\it average} and not for {\it any} state of the system. Obviously, for a system with a translational invariance one only needs a \textit{single} spherical field equation, i.e., $\lambda_{\bf r}=\lambda$ for all ${\bf r} \in {\cal L}$.

\subsection{A short definition of the imperfect Bose gas}
\label{sec:IBG}

Let us consider in a bit more microscopic details a model of an interacting Bose gas. We will only deal with a such model in which the repulsive pair interaction between identical bosons is described by associating
with each pair of particles some mean energy $(a/V)$, where $a>0$, and $V$ denotes the volume
occupied by the system. The Hamiltonian of such imperfect Bose gas \cite{Da72} composed of $N$ particles is defined as 
\begin{equation}
\label{HMF}
H  = H_{0} + H_{mf},
\end{equation}
i.e., is the  sum of the kinetic energy 
\begin{equation}
\label{HPG}
H_{0} = \sum_{\bf{k}}\frac{\hbar^2 {\bf k}^{2}}{2m}{\hat n}_{\bf{k}},
\end{equation}
and the term representing the mean-field approximation to the interparticle interaction
\begin{equation}
\label{mf}
H_{mf} = \frac{a }{V}\frac{N^2}{2 }. 
\end{equation}
The symbols $\{{\hat{n}}_{\bf{k}}\}$  denote the particle number operators and the summation is over one-particle states $\{\bf{k}\}$.

\subsection{Results for the scaling function of the excess free energy of the spherical model}
\label{sec:SM_excess}

The results for that case have been reported in Ref. \cite{D96} and Ref. \cite{D98}. There, the behavior of the model is investigated as a function both on temperature and the magnetic field. Here we will be interested only in its temperature behavior. Then, for the scaling function of the excess free energy one reports
 \begin{eqnarray}
 \lefteqn{X^{({\rm SM})}_{\rm {ex}}(x_\tau) = -\frac 1{2\pi }\Bigg[ \frac 16\left(
 y_L^{3/2}-y_\infty ^{3/2}\right) +\sqrt{y_L}\;{\rm {Li}_2}\left(e^{-\sqrt{y_L}}\right)}\nonumber \\&&   +{\rm {Li}_3}\left( \exp \left( -\sqrt{y_L}
 \right) \right) \Bigg ] -\frac{1}{8\pi}
 x_\tau\left( y_\infty -y_L\right),  \;\;\;\;\;\;\;\;\;\;\;\;\;\;\;\;\;\;\;\;\;\;\;
 \label{xecff} 
\end{eqnarray}
 where 
 \begin{equation}
  \label{sv_tau_SM}
  x_\tau=4\pi K_c \tau L.
  \end{equation}
  For a system with isotropic short-ranged interaction $J$ the critical coupling $K_c$, where $K=\beta J$, has been shown in Ref. \cite{JZ2001} to be 
  \begin{equation}
  \label{JCart}
  K_c=\frac{\left(\sqrt{3}-1\right) \Gamma \left(1/24\right)^2 \Gamma
  	\left(11/24\right)^2}{192 \pi ^3}\backsimeq 0.252731.
  \end{equation}
 In \eqref{xecff} $y_L\equiv y_L(x_\tau)$ and $y_\infty\equiv y_\infty(x_\tau)$ are to be determined from the equations 
 \begin{equation}
  -x_\tau=-2\ln \left[ 2\sinh \left( \frac 12\sqrt{y_L}\right) \right] ,  \label{eqffSM}
  \end{equation}
  for the finite system, and 
  \begin{equation}
  x_\tau=\sqrt{y_\infty } 
  \label{eqifSM}
  \end{equation} 
 for the infinite one,  when $x_\tau\ge 0$. When  $x_\tau<0$, one has $y_\infty=0$. 
 
Making the identifications   
\be
\label{eq:transformation}
\frac{1}{2}\sqrt{y_L^{\left(\rm SM\right)}}=y_L^{\left(\rm Bose \right)}, \quad \frac{1}{2}\sqrt{y_\infty^{\left(\rm SM\right)}}=y_\infty ^{\left(\rm Bose\right)},
\ee
one concludes that
\be
2X_{{\rm ex}}^{\left(\rm SM \right)}=X_{{\rm ex}}^{\left(\rm Bose\right)}
\label{eq:SM_Bose_app}
\ee
The last means that the relativistic Bose gas is mathematically equivalent  to the "two component" spherical model. We recall the same is true also for the imperfect Bose gas \cite{NJN2013,DiRu2017}. Below we demonstrate that. 

\subsection{Results for the scaling function of the free energy of the imperfect Bose gas}
\label{sec:IBG_free_energy}

Below we present some explicit expressions for the model of the imperfect Bose gas reported in \cite{NJN2013}. 
For $d\in (2,4)$ and below the bulk condensation temperature $T<T_c(\mu)$ and $\hat{\mu}\ge 0$ the scaling function $X_{{\rm ex}, {\rm IB}}(x|d)$ takes the form 

\begin{eqnarray}
\label{eq:ex_f_e_B_int}
-X_{\rm ex}^{\rm IB}(x|d)&=& \frac{\zeta(d/2)}{4 \pi} x \left[\sigma(x)\right]^2 + \frac{\Gamma(-d/2)}{2^d \pi^{d/2}}  \left[\sigma(x)\right]^d \nonumber \\
&& + 
\frac{2^{2-d/2}}{\pi^{d/2}} \sum_{n=1}^{\infty} \, \left[\frac{\sigma(x)}{n}\right]^{d/2} K_{d/2}\left[n\,\sigma(x)\right],
\end{eqnarray}
with $\sigma(x)$ obtained as a solution of
\begin{eqnarray}
\label{eq:sigma_eq}
&&x \,\zeta\left(\frac{d}{2}\right)\pi^{d/2-1}-\frac{\Gamma(1-d/2)}{2^{d-2}}\left[\sigma(x)\right]^{d-2}\nonumber \\&&=
2^{3-d/2}\left[\sigma(x)\right]^{d/2-1}\sum_{n=1}^{\infty}n^{-(d/2-1)}K_{d/2-1}[n\sigma(x)]\;.
\end{eqnarray}

Here
\be
\label{eq:def_var_Imp_Bose}
x=\hat{\mu}\left(L/\lambda\right)^{d-2}, \quad \hat{\mu}=(\mu-\mu_c)/\mu_c,
\ee
with
\begin{equation}
\mu_c(T)={\rm Li}_{d/2}(1)\left[a/\lambda^d\right] =\zeta\left(d/2\right)\left[a/\lambda^d\right].
\end{equation}
Performing the identifications
\be
\label{eq:identif_Imp_B_SM}
\sigma =\sqrt{y_L}, \quad \mbox{and} \quad -x_\tau= x\, \zeta \left(d/2\right),
\ee
after setting $d=3$, one obtains, compare with \eq{eq:SM_Bose}, that 
\be
\label{eq:rel_Imp_Bose_SM_app}
X_{\rm ex}^{(\rm IB)}=2 X_{\rm ex}^{(\rm SM)}.
\ee
The last again means that the relativistic Bose gas is mathematically equivalent  to the "two component" spherical model.


\providecommand{\newblock}{}

\end{document}